\documentclass[prl,aps,preprint,showpacs,superscriptaddress]{revtex4}

\usepackage{graphicx}

\begin{document}

\title{Ordered Semiconducting Nitrogen-Graphene Alloys}

\author{H. J. Xiang} 
\affiliation{Key Laboratory of Computational Physical Sciences (Ministry of Education), and Department of Physics, Fudan
  University, Shanghai 200433, P. R. China}

\author{B. Huang}
\affiliation{National Renewable Energy Laboratory, Golden, Colorado 80401, USA}

\author{Z. Y. Li}
\affiliation{Hefei National Laboratory for Physical Sciences at
  Microscale, University of Science and Technology of China, Hefei, Anhui
  230026, P. R. China}

\author{S.-H. Wei}
\affiliation{National Renewable Energy Laboratory, Golden, Colorado 80401, USA}

\author{J. L. Yang}
\affiliation{Hefei National Laboratory for Physical Sciences at
  Microscale, University of Science and Technology of China, Hefei,
  Anhui 230026, P. R. China}

\author{X. G. Gong}
\affiliation{Key Laboratory of Computational Physical Sciences
  (Ministry of Education), and Department of Physics, Fudan
  University, Shanghai 200433, P. R. China}

\date{\today}


\begin{abstract}
The interaction between substitutional nitrogen atoms in graphene is
studied by performing first principles calculations. The nearest neighbor
interaction between nitrogen dopants is highly repulsive because of
the strong electrostatic repulsion between nitrogen
atoms, which prevents the full phase separation in nitrogen doped
graphene.  Interestingly, there are two relatively stable
nitrogen-nitrogen pairs due to the anisotropy charge
redistribution induced by nitrogen doping. We reveal two stable semiconducting ordered N 
doped graphene structures C$_3$N and C$_{12}$N through the cluster
expansion technique and particle swarm optimization method. In particular, C$_{12}$N has a direct band gap of 0.98
eV. The heterojunctions between C$_{12}$N and graphene nanoribbons might
be promising organic solar cells.

\end{abstract}


\maketitle

\section*{Introduction}
Graphene, a single layer of carbon atoms arranged in a honeycomb
lattice, has been the focus of recent research efforts
\cite{Novoselov2005,Novoselov2006,Zhang2005}, due to its unique
zero-gap electronic
structure and the massless Dirac fermion behavior. The unusual
electronic  and structural properties make graphene a promising
material for the next
generation of faster and smaller electronic devices.
However, graphene lacks 
an essential feature for controlled and reliable transistor operation, 
namely, a band gap around the Fermi level. 
Several schemes have been proposed to open a band gap in graphene.
One is the substrate-induced gap for graphene supported on SiC \cite{Zhou2007},
but the experimental realization of this idea turned out  to be very
difficult and controversial \cite{Rotenberg2008}.
Another approach is
the creation of gaps through confinement \cite{Yang2007}, such as in
narrow graphene
nanoribbons (GNRs). 
However, a large scale production of narrow GNRs
is still very challenging.
In addition, it was demonstrated that the band gap of a graphene bilayer can be
controlled externally by applying a gate bias
\cite{Zhang2009,Castro2007}.
Unfortunately, bandgap tailoring by external electrostatic
gate has limited tunability  for the transistor on-off ratio.  
Chemical functionalization \cite{Bekyarova2009,Sarkar2011} is an alternative way to manipulate the
electronic properties of graphene. 
The hydrogenation of
graphene \cite{Sofo2007,Bekyarova2009,Ryu2008,Boukhvalov2009,Xiang2009,Xiang2010b,Lu2009,Zhou2009},
as a prototype of covalent chemical 
functionalization, was studied extensively. 
However, the fully hydrogenated graphene, i.e.,
graphane CH, is an insulator with a very large band gap (larger than 5
eV) \cite{Lebegue2009},
which prevents from using graphane in a field-effect
transistor. There were some efforts to tune the band gap of
two-dimensional (2D) graphene by varying the hydrogen
concentration. Unfortunately, the phase
separation between bare graphene and graphane will take place
spontaneously, leading to an essential zero band gap for the whole
system \cite{Xiang2009}. Similar phenomenon was found to also occur in the 
oxidized graphene \cite{Xiang2010}.

Previous studies \cite{Yu2010,Li2009,Biel2009,Lherbier2008,Huang2011,Wang2009} revealed that doping graphene
and related materials with nitrogen is effective
to tailor its electronic property and chemical reactivity because of
the stronger electronegativity of nitrogen compared
to that of carbon and conjugation between the nitrogen lone pair
electrons and the graphene $\pi$-system. This could create novel
nanomaterials and expand its already widely explored potential
applications. Experiments \cite{Deng2011} suggest that nitrogen species have been incorporated
into the graphene structure with content in the range of
4.5-16.4\%. 
Usually, the nitrogen atoms are substitutionally
incorporated into the basal plane of graphene in the forms of
pyridinic, pyrolic and graphitic nitrogen bonding
configurations.
X-ray photoelectron spectroscopy showed that the pyridinic and pyrolic
nitrogen may lie at the edge or the defect sites and the graphitic
nitrogen (the nitrogen replacing carbon in graphene
plane) is dominant \cite{Liu2011}. A recent scanning tunneling microscopy study
also showed that individual nitrogen atoms were
incorporated as graphitic dopants \cite{Zhao2011}.
Theoretically, the isolated substitutional nitrogen doping in graphene
was studied before \cite{Zhao2011}. 
However, the nature of the interaction between substitutional nitrogen
atoms is far from clear. How the electronic structures of graphene is
modified by multiple nitrogen dopants is an interesting open question to be
addressed.

In this paper, we
examine the collective behavior of nitrogen doping in
graphene on the basis of density functional calculations.  
We find that the nearest neighbor
interaction between nitrogen dopants is highly repulsive because of
the strong electrostatic repulsion between negatively charged nitrogen
atoms.  Our calculations reveal two stable ordered N 
doped graphene structures C$_3$N and C$_{12}$N. Surprisingly, nitrogen
atoms do not simply doping electrons to the graphene $\pi^*$
state. Instead, both ordered structures are found to be
semiconducting. In particular, C$_{12}$N has a direct band gap of 0.98
eV. The substitutional nitrogen raises the valence band maximum (VBM)
and conduction band minimum (CBM) level in N
doped graphene, leading to  a  type-II band alignment between C$_{12}$N and 6-AGNR.
We propose that C$_{12}$N/6-AGNR is a promising  
organic solar cell with C$_{12}$N as donor and 6-AGNR as acceptor,
respectively.

\section*{Computational Method}
In our DFT calculations,
the local density approximation (LDA) was adopted unless otherwise stated. 
The plane-wave cutoff energy for wavefunction was set to 500 eV.
The ion-electron interaction was treated using the projector
augmented wave (PAW) \cite{PAW} technique as implemented in the Vienna
ab initio simulation package \cite{VASP}. 
For the Brillouin zone sampling, we use $2\times2\times1$ Monkhorst-Pack k-mesh
for the $12\times12\times1$ graphene supercell. And for C$_3$N (C$_{12}$N), we used the
$12\times12\times1$   ($6\times6\times1$ ) Monkhorst-Pack k-mesh. The
constant temperature first-principles molecular dynamics simulations
were performed to check the 
thermal stability of C$_3$N and C$_{12}$N.
The ``cluster expansion'' \cite{Ferreira1989}
of the alloy Hamiltonian was carried out by the ATAT package
\cite{atat}.   
The global minimum optimization was performed using the
two-dimensional (2D) particle swarm optimization (PSO) technique, which we
proposed recently to predict the most stable 2D crystals \cite{Luo2011}.
For the PSO simulations, we 
use the Crystal structure AnaLYsis by Particle Swarm
Optimization (CALYPSO) code \cite{Wang2010}.

\section*{Results and Discussion}
\subsection*{Interaction Between Nitrogen Dopants}
\begin{figure}
    \begin{center}
    \includegraphics[width=10.5cm]{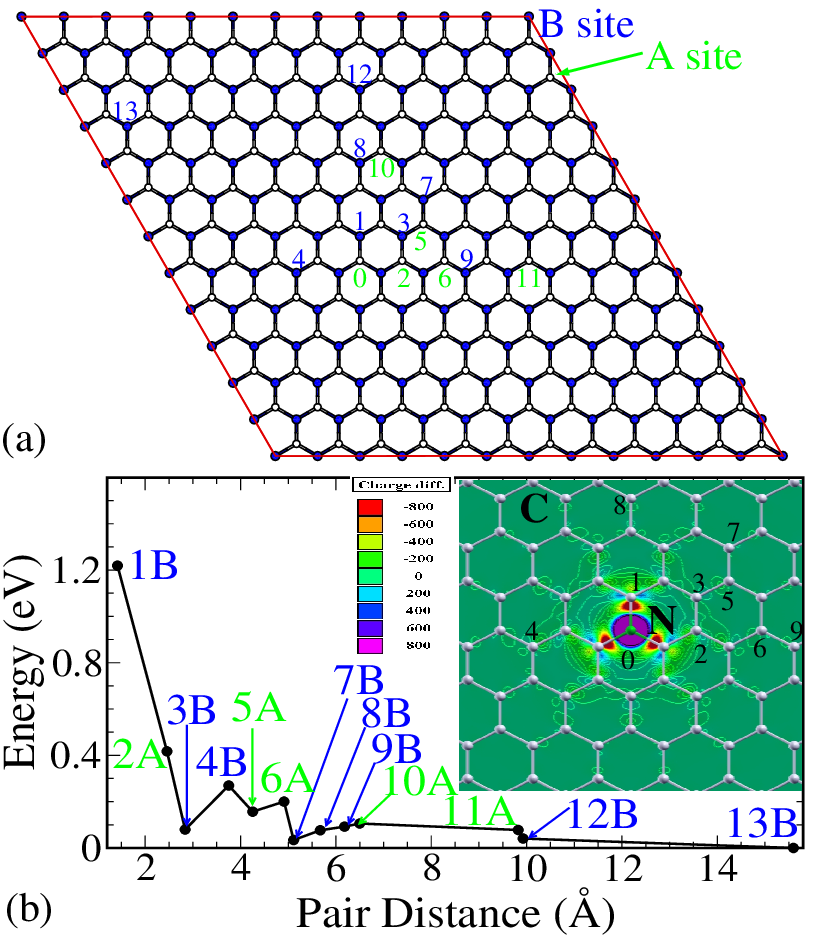}
    \caption{ (a) The $12\times12$ graphene supercell. The solid or
      empty spheres belong to sublattice A or B, respectively. The numbers
      label the carbon lattice sites for nitrogen substitution. (b)
      The pair interaction energy between nitrogen atoms as a function
      of the pair distance from the DFT
      total energy calculations for the relaxed structures. For each N-N
      pair, one of the nitrogen atom is at 0A site and the other
      nitrogen atom is labeled in (a). The inset shows the contour
      plot of the electron charge
      density difference between the isolated N doped graphene and
      undoped graphene.}
    \label{fig1}
    \end{center}
  \end{figure}

To investigate the interaction between two substitutional nitrogen
atoms, we use a $12\times12$ supercell of graphene. The use of
such a large supercell is essential due to the relative long range
interaction between nitrogen dopants. And test calculation shows that
using larger supercells gives similar results.
As is well known, 
the honeycomb lattice of graphene consists of two interpenetrating
triangular sublattices, namely sublattice A and sublattice B. We
consider the interaction between a nitrogen atom at 0A site and those at other
sites [see Fig.~\ref{fig1}(a)]: All interaction with the interaction range less than 6.5
\AA\ are included, and we also include several other longer range
interaction N-N pairs (0A-11A, 0A-12B, 0A-13B). For each N-N pair, 
the internal atomic structure is fully relaxed. With respect to the
energy of longest N-N pairs 0A-13B considered in this work (15.60 \AA), We plot the
interaction energy of other N-N pairs in Fig.~\ref{fig1}(b). First, we
can see that all the interaction energy are positive, i.e., the
interactions are repulsive. This can be understood because nitrogen has a
high electronegativity than carbon thus nitrogen atoms are surrounded
by more electrons which leads to the repulsive electrostatic
interaction. Second, the nearest neighbor interaction is extremely
repulsive. This is because the two nitrogen atoms
are directly connected along the bond direction leading to a very strong electrostatic repulsion.
Third, generally the interaction energy decreases with
the pair distance because of the nature of coulombic
electrostatic interaction.
Fourth, for N-N pairs with short pair distances (less than 6.5
\AA), the interaction energy curve has two local minima at N-N pairs
0A-3B and 0A-7B. The interaction between 0A and 3B and
that between 0A and 7B are only slightly repulsive with the
interaction energies of 0.08 eV and 0.03 eV, respectively. 

The nontrivial non-monotonic dependence of the pair interaction energy on the
N-N distance and the presence of local minima at N-N pairs
0A-3B and 0A-7B are interesting findings to be understood. 
Our calculations show that the strain relaxation is not responsible
for the stability of the 0A-3B and 0A-7B  N-N pairs because the
energies of the unrelaxed pair structures display the same trend. 
Instead, we find that the stability of the 0A-3B and 0A-7B  N-N pairs is due to the
anisotropic electron charge density redistribution induced by the nitrogen
substitution. Fig.~\ref{fig1}(b) shows the charge difference between the
isolated N-doped graphene and the undoped graphene. We can see that
the charge redistribution in graphene is mainly along the nearest neighbor 
bonding direction. Since sites 3B and 7B are nearly opposite to the bonding
direction, there is  much less charge redistribution around site 3B (7B)
than that around 2A and 4B (6A and 8B). Therefore, in the case of the 0A-3B and
0A-7B  N-N pairs, the coulombic repulsion will be weaker than that between the
other neighboring pairs.

Our calculations show that the interaction between substitutional nitrogen atoms in
graphene differs fundamentally from that between adsorbed hydrogen (oxygen) atoms on
graphene. In the hydrogenated (oxidized) graphene case,
hydrogen (oxygen) tends to be close to each other in order to
lower the kenetic energy of the carbon $\pi$ electrons,
leading to the phase separation between fully hydrogenated (oxidized) graphene
and bare graphene\cite{Xiang2009,Xiang2010}.
The repulsive interaction between substitutional nitrogen atoms will prevent the phase separation of
nitrogen doped graphene into undoped graphene part and highly nitrogen
doped graphene part, in contrast to the case of the hydrogenation and
oxidization of graphene \cite{Xiang2009,Xiang2010}.

\subsection*{Ordered Semiconducting Nitrogen-Graphene Alloys}
\begin{figure}
    \begin{center}
    \includegraphics[width=10.5cm]{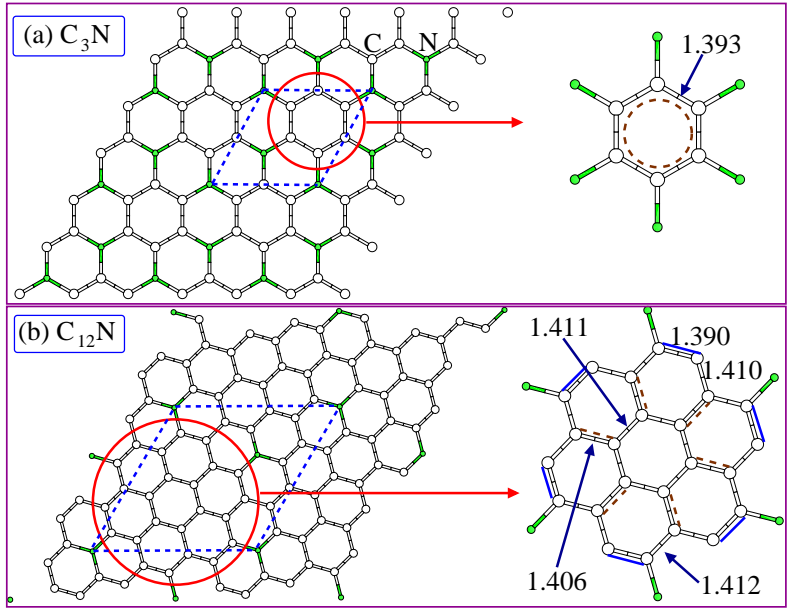}
    \caption{ (a) shows the ordered N doped graphene superstructure
      C$_3$N. (b) shows the ordered N doped graphene superstructure
      C$_{12}$N. The unit cells of the superstructures are enclosed by
      the dashed lines. 
      The right panels displays the enlarged view of the
      structures enclosed by the circles of the left panels. The
      numbers denote the bond lengths (in \AA). The Clar formulas are
      also shown. }
    \label{fig2}
    \end{center}
  \end{figure}

In our above discussions, 0A-3B and 0A-7B are two relatively
stable configurations of N-N pairs. For each type of two pairs, we can
construct a graphene superstructure with a uniform distribution of
nitrogen dopants. For the 0A-3B N-N pair, the ordered nitrogen doped graphene
structure is a $2\times 2 $ superstructure (C$_3$N) of the
graphene, as shown in Fig.~\ref{fig2}(a). We can see that the doped
nitrogen atoms themselves form a honeycomb lattice, similar to 
graphene but with a doubled lattice constant. As a consequence, the
carbon atoms are separated into isolated six-membered rings, as shown
in the right panel of Fig.~\ref{fig2}(a). For the 0A-7B N-N pair, the
ordered nitrogen doped graphene
structure is a $\sqrt{13}\times \sqrt{13}$-R46.1$^\circ$ superstructure (C$_{12}$N) of the
graphene, as shown in Fig.~\ref{fig2}(b). In this superstructure, the
carbon atoms are connected by carbon-carbon bonds to form a single
carbon domain, in contrast to the C$_3$N case. 
 
To examine the stability of the C$_3$N and C$_{12}$N superstructures,
we use  ``cluster expansion'' method \cite{Ferreira1989}  and
direct comparison of total energies between different
superstructures. 
In the ``cluster expansion'' method,  
the alloy Hamiltonian is mapped onto a generalized Ising  Hamiltonian.
In brief, for some nitrogen doped graphene configurations, we perform spin polarized DFT
calculations
to relax the cell and internal atomic coordinates. The energies of
the
relaxed structures are used to extract the interaction parameters of
the alloy  Hamiltonian. 
After obtaining the interaction
parameters (See Supplemental Material \cite{support})
 of the alloy  Hamiltonian, we find that the C$_3$N
superstructure shown in  Fig.~\ref{fig2}(a) is the most stable structure
for the C$_x$N alloy with the 25\% nitrogen concentration (at least
for cells with no more than 32 atoms). For the C$_{12}$N
superstructure with the 0A-7B N-N pairs, the relatively long range
(5.11 \AA) pair interaction is not included in the alloy
Hamiltonian. Therefore, we generate all superstructures with no
more than 26 atoms and the same nitrogen concentration (7.7\%) as
C$_{12}$N  using the linear scaling algorithm recently proposed by
Hart and Forcade\cite{Hart2009}. Our calculations show that the C$_{12}$N superstructure
shown in Fig.~\ref{fig2}(b) has the lowest energy among all considered
structures. The lattice dynamics calculation shows that both C$_3$N
and C$_{12}$N have no imaginary frequency phonon and thus are
stable \cite{support}. 
First-principles molecular dynamic simulations up to 500 K were
performed to check the stability of C$_3$N and C$_{12}$N. We find that the N
atoms only vibrate around the equilibrium position, which confirms the
thermal stability of C$_3$N and C$_{12}$N.

\begin{figure}
    \begin{center}
    \includegraphics[width=10.5cm]{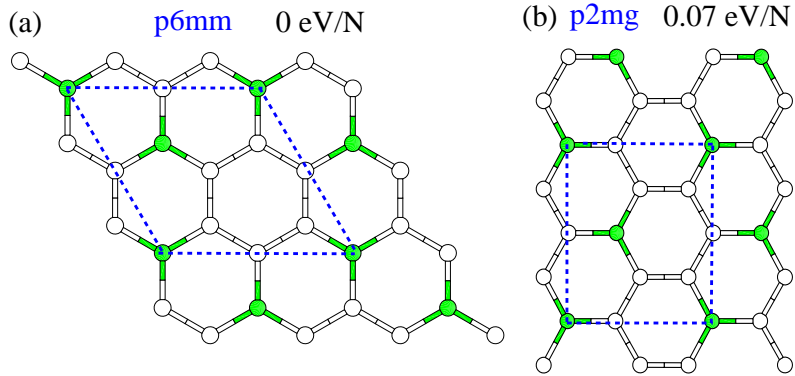}
    \caption{ (a) shows the most stable 2D structure (plane
      group: $p6mm$) of C$_3$N from the 2D PSO simulations. (b) shows
      the second lowest energy 2D structure (plane group: $p2mg$) of
      C$_3$N from the 2D PSO simulations. The unit cells of the
      superstructures are enclosed by
      the dashed lines. The numbers denote the relative energies.} 
    \label{fig3}
    \end{center}
  \end{figure}

Above we proposed two possible stable ordered nitrogen-carbon alloys
(C$_3$N and C$_{12}$N shown in
Fig.~\ref{fig2}) by considering only the graphene-based structures.
Here, we perform
PSO global minimum optimization to confirm the proposed structures. In
the 2D PSO simulations \cite{Luo2011}, the graphene lattice structure is not
assumed. Instead, we generate random structures (both atomic positions
and cell parameters) to initialize the simulations. The initial
structures are subsequently relaxed before performing the 
PSO operations. The population size is set to 30. We consider
all possible cell sizes with the total number of atoms no more than
26. The number of generations is fixed to 30. Our simulations show
that the most stable 2D nitrogen-carbon alloy structure with 25\% N
concentration is indeed the structure ($p6mm$ plane group) shown in Fig.~\ref{fig2}(a) [see
also Fig.~\ref{fig3}(a)]. The second lowest energy 2D structure with
the plane group $p2mg$ [Fig.~\ref{fig3}(b)] has a higher energy by 0.07 eV/N.
It should be noted that both $p2mg$ and $p6mm$ structures can be
viewed as nitrogen doped graphene structures although the graphene
honeycomb lattice is not assumed in the PSO simulations. 
The PSO simulations for the 2D nitrogen-carbon alloy structure with
7.7\% nitrogen concentration do not give structures more stable than
the C$_{12}$N shown in Fig.~\ref{fig2}(b), in support of the stability
of the proposed C$_{12}$N  structure. We note that the formation
energies of C$_3$N and C$_{12}$N are positive if graphene and N$_2$ molecule are
taken as the references. However, many experiments
\cite{Wang2009,Zhao2011} showed that N atoms
can be incorporated into graphene. Here we show that C$_3$N and C$_{12}$N are
the most stable structures once N atoms are incorporated into
graphene, as can be seen from the convex hull plot in the new
supporting material \cite{support}.

\begin{figure}
    \begin{center}
    \includegraphics[width=10.5cm]{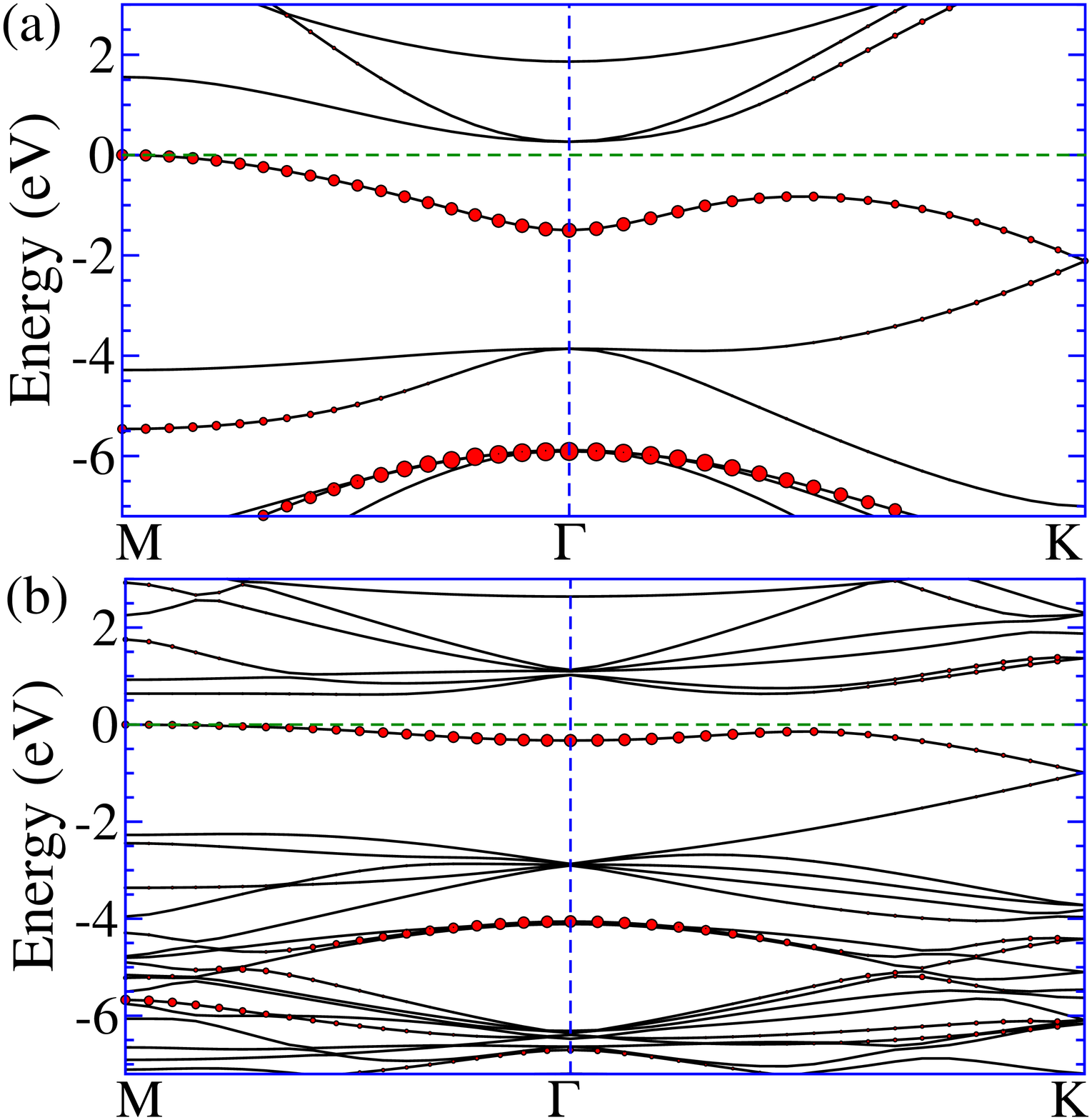}
    \caption{ The LDA band structures for (a) C$_3$N and (b)
      C$_{12}$N. The horizontal dashed lines denote the VBM
      levels. The contributions from N $p_z$ orbitals to the
      wavefunctions are indicated by
      the radii of solid spheres. }
    \label{fig4}
    \end{center}
  \end{figure}

The band structures for the C$_3$N and C$_{12}$N superstructures from
the LDA calculations are shown in
Fig.~\ref{fig4}(a) and (b), respectively. Interestingly, both superstructures are found
to be semiconductors: C$_3$N has an indirect band gap (about 0.26 eV
from the LDA calculation) with the VBM at $M$ and CBM at $\Gamma$,
respectively; C$_{12}$N has instead a direct band gap (about
0.62 eV from the LDA calculation) at 
the $M$ point.
Because the $p_z$ orbital of a nitrogen atom will be
almost fully occupied by the two $\pi$ electrons,
the mechanism for the band gap opening in graphene
induced by nitrogen doping can
be understood by examining only the bonding situation of carbon $\pi$
electrons. Recently, Clar's
theory of the aromatic sextet \cite{Clar1972} was shown to be a simple and powerful tool to predict
the stability and the electronic/magnetic structure of graphene
related systems. According to Clar's rule, for a given system, the representation with a
maximum number of Clar sextets, called the Clar formula, is
the most representative one. Here, we show the Clar formulas for C$_3$N
and C$_{12}$N in the right panels of Figs.~\ref{fig2}(a) and (b),
respectively.
In C$_3$N, each six-membered ring forms a Clar sextet. And C$_3$N has one unique Clar formula, thus belongs
to the class 1CF of the (pseudo)-all-benzenoid structure. In contrast, C$_{12}$N
has no Clar sextets, and thus are nonbenzenoid. Because each nitrogen
atom forms three single bonds with the neighboring carbon atoms, the
nitrogen atoms are excluded when constructing the Clar formula. 
We note that in the Clar formula for the -NH terminated graphene
nanoribbon, the two N-C bonds were also treated as single bonds \cite{Seitsonen2010}.
The Clar formulas
shown in Figs.~\ref{fig2} are in good agreement with the C-C bond lengths
of relaxed structures from the DFT calculation. All the  C-C bond lengths in  C$_3$N are the
same, i.e., 1.393 \AA. For C$_{12}$N, some C-C bonds (marked by an
additional solid line) have short bond lengths, 1.388 \AA, which can
be considered as normal C-C double bond; and there are weak C-C double
bonds (marked by an additional dashed line) with the bond lengths of
1.406 \AA; all the other C-C bonds are single bond. 
Wassmann {\it et al.} \cite{Wassmann2010}
showed that all graphene related structures has a relative large band
gap except for the (pseudo)-all-benzenoid structure with class nCF
($n>=3$) such as graphene which has small or zero band gap. Therefore,
we expect that there is a large band gap between the occupied carbon
$\pi$ states and the unoccupied carbon $\pi$ states in both C$_3$N and
C$_{12}$N.  Our analysis for the wavefunction characters indicates
that the highest occupied band near $\Gamma$ is mainly contributed by nitrogen $p_z$
orbitals and there are some contributions from carbon
$p_z$ orbitals when approaching the Brillouin zone boundary. The lowest
unoccupied band is mostly contributed by the carbon $p_z$ orbitals.
The gap between the highest occupied carbon  $\pi$ state and the CBM is
larger than 1.5 eV (in LDA) in both C$_3$N and
C$_{12}$N, which is in accord with the Clar's theory.   The small band gaps
for  C$_3$N and C$_{12}$N are due to the presence of the nitrogen $p_z$
state (VBM state) above the highest occupied carbon $\pi$ state. 
C$_3$N has a smaller band gap because the direct hopping
between N $p_z$ orbitals in C$_3$N leads to the larger dispersion
of the highest occupied band. We note that the dilute limit of
nitrogen substitutional doping in graphene was experimentally studied 
by Zhao {\it et al.} \cite{Zhao2011}, where 
the extra electron of the nitrogen dopant
was found to delocalized into the neighboring graphene lattice,
resulting in the n-type behavior. In this work, we predict that
collective nitrogen doping can induce exotic semiconducting behavior
in graphene. Experimentally, some disorder in the nitrogen
distribution might occur due to the thermal fluctuation. To see the
effect of the disorder on the band gap, we have simulated a disordered
configuration with 12.5\% N concentration using a 128-atom
special quasirandom structure \cite{SQS}. The LDA band
gap is found to be 0.08 eV. The real band gap
should be larger because the LDA functional
underestimates the band gap. This suggests
that the disorder might change the magnitude
the band gap, but shall not close
it. Moreover, because the ordered C$_3$N and
C$_{12}$N is more stable at low temperature than
the disordered one, we believe reasonably
large band gap should exist in nitrogen doped
graphene synthesized at low temperature.

It is well-known that LDA seriously
underestimates the band gap of semiconductors, thus we calculate the
electronic structures of C$_3$N and C$_{12}$N by employing the
screened Heyd-Scuseria-Ernzerhof 06 (HSE06) hybrid functional
\cite{Heyd2003,Krukau2006,Paier2006},  which was shown 
to give a good band gap for many semiconductors including graphene
related $\pi$ systems \cite{Hod2008}. Our HSE06 calculations show that
C$_3$N and C$_{12}$N have the band gaps of 0.96 eV and 0.98 eV,
respectively. It is noted that the band gaps of the
nitrogen doped graphene are close to the band gap (about
1.12 eV) in silicon. This suggests that C$_3$N and C$_{12}$N with
similar electronic properties as silicon might be promising electronic
materials for the next generation CMOS technology.   
 
\subsection*{Potential Organic Solar Cell}
\begin{figure}
    \begin{center}
    \includegraphics[width=10.5cm]{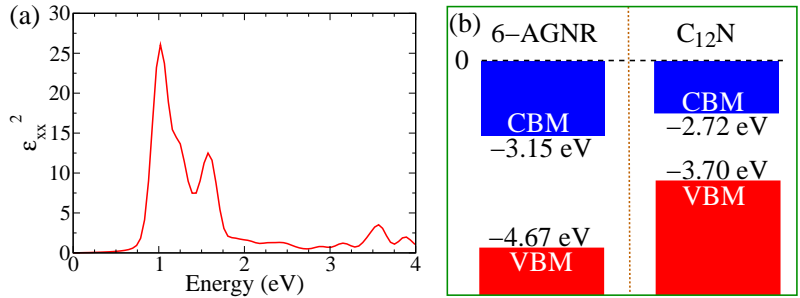}
    \caption{ (a) shows the computed imaginary part
      of the frequency dependent dielectric function from the screened
      exchange HSE06 calculation. (b) illustrates the band offset between C$_{12}$N and
  6-AGNR from the HSE06 calculations. The numbers are the VBM and CBM
  levels with respect to the vacuum level.}
    \label{fig5}
    \end{center}
  \end{figure}

The fact that C$_{12}$N has a direct band gap of 0.98
eV suggests that C$_{12}$N might be a promising solar cell absorption
material. To investigate this possibility, we calculate the imaginary part
of the frequency dependent dielectric function via summation over
pairs of occupied and empty states without considering the local field effects using the
HSE06 screened hybrid function.
Fig.~\ref{fig5}(a) shows the calculated result ($\epsilon_{xx}^2$) for one of the diagonal
parts of the in-plane components.   We can see that there is a very
strong peak at around 1.0 eV arising from the interband HOMO-LUMO
transition. Such strong adsorption peak is due to the fact that both
the VBM band and CBM band around M are rather flat. Due to the
dispersion of the VBM band and CBM band at other K-points, there is
also substantial optical absorption up to 1.6 eV. The absorption
between 1.9 eV and 4.0 eV is rather weak, indicating that a solar cell
with only C$_{12}$N might be not efficient because the solar energy
between 1.9 eV and 4.0 eV may not be absorbed completely. Our HSE06
calculation shows that C$_{12}$N has a relatively high VBM level
($-3.70$ eV with repect to the vacuum level). This suggests that
C$_{12}$N can act as a donor. If an appropriate acceptor
material can be found, we can construct a photovoltaic system with a similar
architecture as the P3HT/C$_{60}$-PCBM bulk heterojunction solar cell.
We find from HSE06 calculations that the the VBM and CBM levels of
6-AGNR are $-4.67$ eV and $-3.15$ eV, respectively. Therefore, the
band alignment between C$_{12}$N and 6-AGNR is of type-II, as shown in
Fig.~\ref{fig5}(b). In this proposed C$_{12}$N/6-AGNR solar cell,
C$_{12}$N can absorb the solar energy between 0.98 eV and $\sim$1.50 eV. And
6-AGNR  can absorb the solar energy above the band gap 1.52 eV. The
eletron-hole pair excited by photo can be separated by the field due
to the type-II band alignment.

\section*{Conclusion}
In summary, we have studied the interaction between substitutional
nitrogen atoms in graphene and find that the nearest neighbor
interaction between nitrogen dopants is highly repulsive because of
the strong electrostatic repulsion between nitrogen
atoms.  Our calculations reveal two stable ordered N 
doped graphene structures C$_3$N and C$_{12}$N. Both structures are
semiconducting, in contrast to the common belief that nitrogen atoms
simply dope electrons to the graphene Dirac cone.
In particular, C$_{12}$N has a direct band gap of 0.98 
eV. The substitutional nitrogen raises the VBM
and CBM level in N
doped graphene, leading to  a  type-II band alignment between C$_{12}$N and 6-AGNR.
We propose that C$_{12}$N/6-AGNR is a promising  
organic solar cell with C$_{12}$N as donor and 6-AGNR as acceptor,
respectively. 
Our study suggests that nitrogen doping might be a
promising way to open a band gap in grapene for the application in
electronics and photovoltaics.

\section*{Acknowledgment}
Work at Fudan was partially supported by the National Science Foundation
of China, Pujiang plan, and The Program for Professor of Special Appointment
(Eastern Scholar) at Shanghai Institutions of Higher Learning.
Work at NREL was supported by the LDRD program funded by the U.S. Department of Energy, under
Contract No. DE-AC36-08GO28308.

  \clearpage

\end{document}